\documentclass[%
preprint,
nofootinbib,
 amsmath,amssymb,
 aps,
]{revtex4-2}

\usepackage{graphicx}
\usepackage{dcolumn}
\usepackage{bm}

\begin{document}

\def\bb    #1{\hbox{\boldmath${#1}$}}

\title{The imprints of QCD cascades in hadron multiplicity distribution} 

\author{Maciej Rybczy\'nski}
\email{maciej.rybczynski@ujk.edu.pl}
\affiliation{Institute of Physics, Jan Kochanowski University, 25-406 Kielce, Poland}
\author{Zbigniew W\l odarczyk}
\email{zbigniew.wlodarczyk@ujk.edu.pl}
\affiliation{Institute of Physics, Jan Kochanowski University, 25-406 Kielce, Poland}

\begin{abstract}

The relation between parton and hadron multiplicity distributions is discussed. To obtain parton multiplicity distribution we propose decomposition of the multiplicity distributions of final state hadrons. Such procedure offers hope for experimental testing of physical probes of quantum complexity.

\end{abstract}

\pacs{XXX}

\maketitle

\section{Introduction}
\label{Introduction}

In recent years, quantum information—encompassing concepts such as entanglement and complexity—has offered a novel perspective on high-energy physics. In a series of works (see Ref.~\cite{Caputa} and references therein), it has been argued that understanding entanglement in Quantum Chromodynamics (QCD) could shed new light on the gluon-dominated structure and properties of high-energy hadrons.
In high-energy physics, the investigation of quantum correlations is of paramount importance. Quantum correlations, often manifested as entanglement, play an essential role in describing the behavior of particles and their interactions. These correlations extend beyond the classical domain, enabling particles to exhibit nonlocal connections that challenge conventional notions of reality~\cite{Nadir2024}.

Entanglement, a fundamental phenomenon of quantum mechanics, has evolved from a conceptual puzzle into a central focus of high-energy physics. The principle of superposition, wherein particles can occupy multiple states simultaneously, and entanglement, wherein particles are intrinsically connected across vast distances, form the core of the quantum paradigm~\cite{Aspect}. By transcending classical limits, entanglement challenges traditional views of particle separability. In high-energy physics, this nonlocal interdependence is crucial for understanding phenomena such as confinement, where quarks and gluons remain bound within hadrons~\cite{Wilczek}. Moreover, entanglement entropy not only characterizes the degree of entanglement but also provides insights into the structure of quantum states, revealing connections among particles and their collective behavior.

In the experiments involving multiparticle production, the observable quantities pertain to the final-state hadron distributions, while their connection to the underlying partonic distributions remains an open and unresolved problem.
Within the framework of compound distributions—such as those encompassing the geometric distribution—we propose a method to infer properties of the parton-level dynamics from the experimentally accessible hadronic multiplicity distributions.
The proposed approach offers a novel tool for experimental investigations of quantum complexity in strongly interacting systems.

\section{Parton distribution from toy model of QCD evolution}
\label{QCDevolution}

It will be convenient for us to describe the parton evolution using the dipole representation—in this representation, a set of partons is represented by a set of color dipoles. 
In this model the Balitsky-Fadin-Kuraev-Lipatov (BFKL) equation for the dipole scattering cross section at a rapidity $y$ reproduces the powerlike increase of the cross section with energy, $\exp(\Lambda y)=x^{-\Lambda}$ where $\Lambda$ is the BFKL intercept.

Let us now introduce $P\left(n,y\right)$, which is the probability to find $n$ dipoles (of a fixed size in our model) at rapidity $y$. For this probability we can write the following recurrent equation~\cite{PhysRevD.102.074008,PhysRevD.95.114008}:
\begin{equation}
\frac{dP\left(n,y\right)}{dy}=-\Lambda n P\left(n,y\right)+\Lambda \left(n-1\right)P\left(n-1,y\right).
\label{evolution}
\end{equation}
This equation represents a standard form of a cascade process. The first term accounts for the reduction in the probability of finding $n$ dipoles as a result of their splitting into $n+1$ dipoles, whereas the second term describes the corresponding increase arising from the splitting of $n-1$ dipoles into $n$ dipoles. By employing the generating function method to solve this equation, one obtains a geometric distribution, as detailed in~\cite{Chaturvedi}
\begin{equation}
P\left(n\right) = \Theta\left(1-\Theta\right)^{n-1}, 
 \label{geometric}
\end{equation}
where $\Theta =\exp\left(-\Lambda y\right)$  determine mean multiplicity $\langle n\rangle=\left(1-\Theta\right)/\Theta$. The geometric distribution is the only memoryless discrete probability distribution \footnote{Geometric distribution is the discrete version of the same property as in the exponential distribution. Approximate version of Eq.~(\ref{geometric}) for a continuous variable $z=n/\langle n\rangle$ and the relation to KNO scaling is discused in appendix~\ref{KNO}.}. The resulting Shanon entropy, $S=-\sum_n P\left(n\right)\ln\left[P\left(n\right)\right]$, is
\begin{equation}
S=-\ln \Theta-\frac{1-\Theta}{\Theta}\ln(1-\Theta) = \left(\langle n\rangle+1\right)\ln\left(\langle n\rangle+1\right)-\langle n\rangle\ln\langle n\rangle.
\label{S}
\end{equation}
At large $\Lambda y$ ($x \sim 10^{-3}$) the relation between the entanglement entropy and the gluon structure function $G\left(x\right)=\langle n\left(x\right)\rangle \sim x^{-\Lambda}$ becomes very simple
\begin{equation}
S=\ln\left[G\left(x\right)\right].  
\label{S2}
\end{equation}
It signals that all $\exp\left(\Lambda y\right)$ partonic states have about equal probabilities $\exp\left(-\Lambda y\right)$. In this case the entanglement entropy is maximal, and the proton is a maximally entangled state~\cite{PhysRevD.95.114008}. In the maximally entangled regime at small $x$, it appears that the behavior of the gluon structure function becomes universal. In analogy to statistical mechanics, in thermal equilibrium (maximal entropy), the equation of state is determined by temperature ($1/x$) and the effective number of degrees of freedom.

\section{Decomposition of final state}
\label{decomposition}

Multiplicity distributions, $P\left(N\right)$, where $N$ denotes the number of produced particles, are among the primary observables measured in multiparticle production experiments. They have long been recognized as a crucial source of information regarding 
the underlying dynamics of the particle production mechanism~\cite{Kittel}. The precise relationship between partonic and hadronic multiplicity distributions remains an open question. A common working assumption has been the hypothesis of parton-hadron 
duality~\cite{Dokshitzer}, which posits that the partonic and hadronic distributions are essentially identical. Under this assumption, the hadron multiplicity distribution would be expected to follow the geometric form given in Eq.(\ref{geometric}) \cite{H1,Tu}.
However, this assumption is highly idealized-particularly for hadrons produced in the central rapidity region—since the experimentally measured distributions $P\left(N\right)$ often exhibit significant deviations from the geometric  distribution~(\ref{geometric}).

The negative binomial distribution (NBD) is typically the first choice for fitting multiplicity distributions $P\left(N\right)$ in high-energy collision data. However, as the collision energy and the number of produced secondaries increase, the single NBD 
increasingly fails to accurately describe the data, particularly in the high-multiplicity tail. This discrepancy has motivated the use of more sophisticated models, including two-, three-, or multi-component NBDs, as well as alternative forms of 
$P\left(N\right)$~\cite{RWWPRD}. Since neither a single NBD nor a binomial distribution (BD) provides a satisfactory description of the data across the full multiplicity range, we explore the framework of compound distributions. These are particularly suitable 
in scenarios where the particle production process proceeds via an intermediate stage involving the generation of $M$ sources—such as clusters, fireballs, or other composite objects—distributed according to some function $f\left(M\right)$ with generating function $F\left(z\right)$. Each of these sources subsequently decays independently into a number of secondaries  $n_{ i=1,...,M}$ governed by an identical for all $M$ distribution $g(n)$, characterized by the generating function $G\left(z\right)$. The resultant multiplicity distribution, 
\begin{equation}
h\left(N = \sum_{i=0}^M n_i \right),  
\label{hN}
\end{equation}
is a compound distribution of $f\left(M\right)$ and $ g\left(n\right)$ with generating function
\begin{equation}
H\left(z\right)=F\left(G\left(z\right)\right).  
\label{Hz}
\end{equation}

By summing the contributions from $k$ independent geometric distributions, where each individual multiplicity $n_i \in (0, \infty)$, one obtains the well-known NBD. The corresponding generating function is given by
\begin{equation} 
H\left(z\right) = \left( \frac{\Theta'}{1 - (1 - \Theta')z} \right)^k, 
\label{GNBD} 
\end{equation} 
where $\Theta'$ is a parameter related to the geometric distribution.~\footnote{The geometric distribution in Eq.~(\ref{geometric}) is defined for $n>0$. For the case $n=0,1,2,...$ the parameter transforms as $1/\Theta'=1/\Theta-1$. }

When the number of particle-emitting sources $M$ fluctuates according to a BD with mean $\lambda$ and maximal value $K$, the resulting compound distribution is known as the geometric-binomial distribution~\cite{Eledum}, with generating function: 
\begin{equation} 
H\left(z\right) = \left( 1 + \frac{z - 1}{1 - (1 - \Theta')z} \cdot \frac{\lambda}{K} \right)^K. 
\label{GGB} 
\end{equation}

Alternatively, if the number of sources follows a Poisson distribution with mean $\lambda$, the resulting distribution is referred to as the geometric-Poisson distribution (also known as the P\'olya-Aeppli distribution), whose generating function takes the form: 
\begin{equation} 
H\left(z\right) = \exp\left(\frac{\lambda(z - 1)}{1 - (1 - \Theta)z} \right). 
\label{GGP} 
\end{equation}

Multiplicity distributions $P\left(N\right)$ of the compound geometric distributions (\ref{GNBD}), (\ref{GGB}) and (\ref{GGP}) are given in appendix \ref{PN}.

\section{ Experimental tests }
\label{comparison}

\begin{figure}
\begin{center}
\includegraphics[scale=0.48]{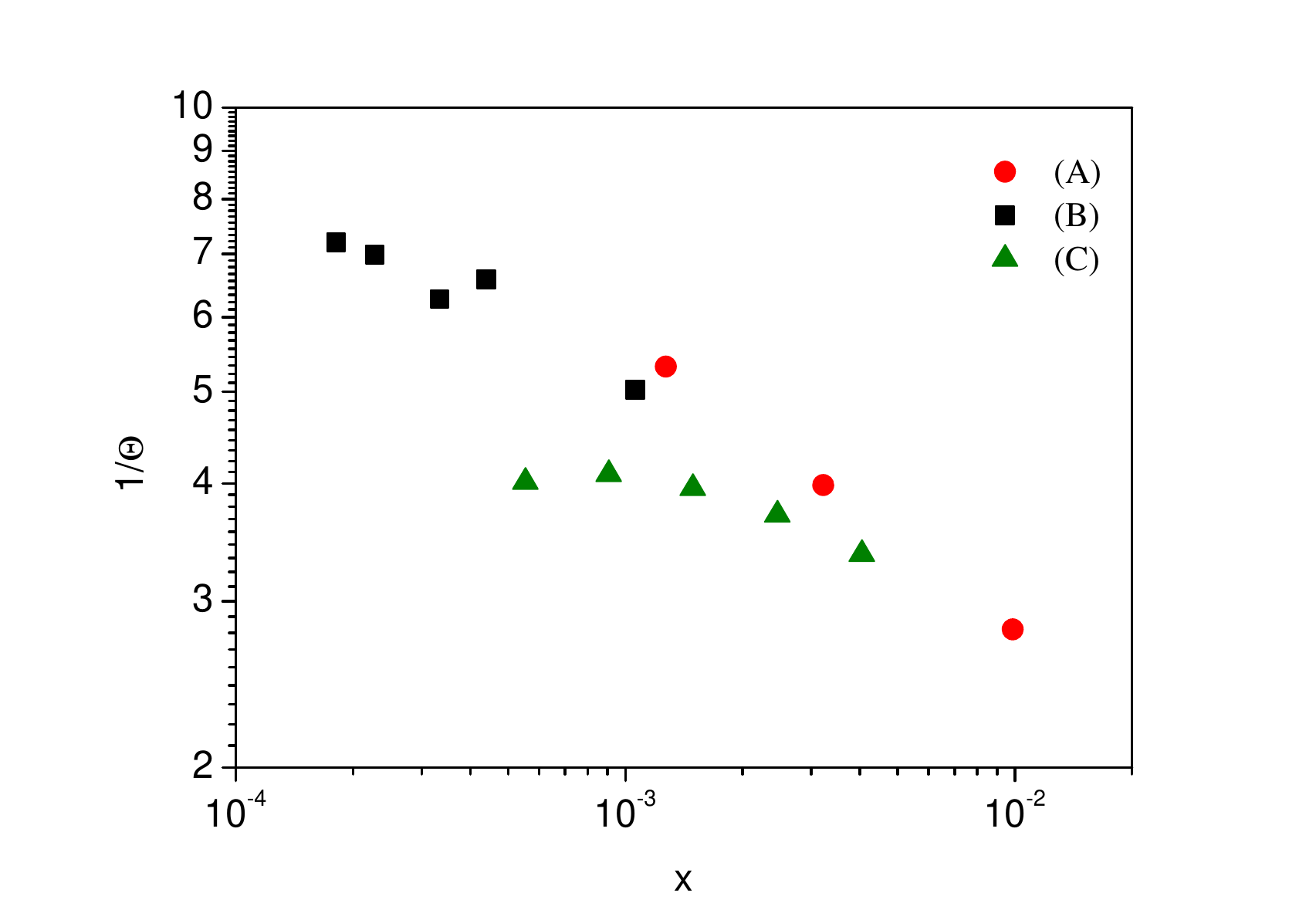}
\end{center}
\vspace{-5mm}
\caption{ Parameter $\Theta$ derived from multiplicity distributions as a function of $x$. (A) circles: $\sqrt s$= 0.9, 2.76 and 7 TeV, $2.3<\eta<3.9$ (photons); (B) squares: $\sqrt s$ = 8 TeV, $|\eta|<2$, $|\eta|<2.4$, $|\eta|<3$,$|\eta|<3.4$ and $-2.4<\eta<5$ (charged particles); (C) triangles: $\sqrt s$= 7 TeV, $2<\eta<2.5$, $2.5<\eta<3$, $3<\eta<3.5$, $3.5<\eta<4$ and $4<\eta<4.5$ (charged particle).}
\label{SPD-1}
\end{figure}

\begin{figure}
\begin{center}
\includegraphics[scale=0.48]{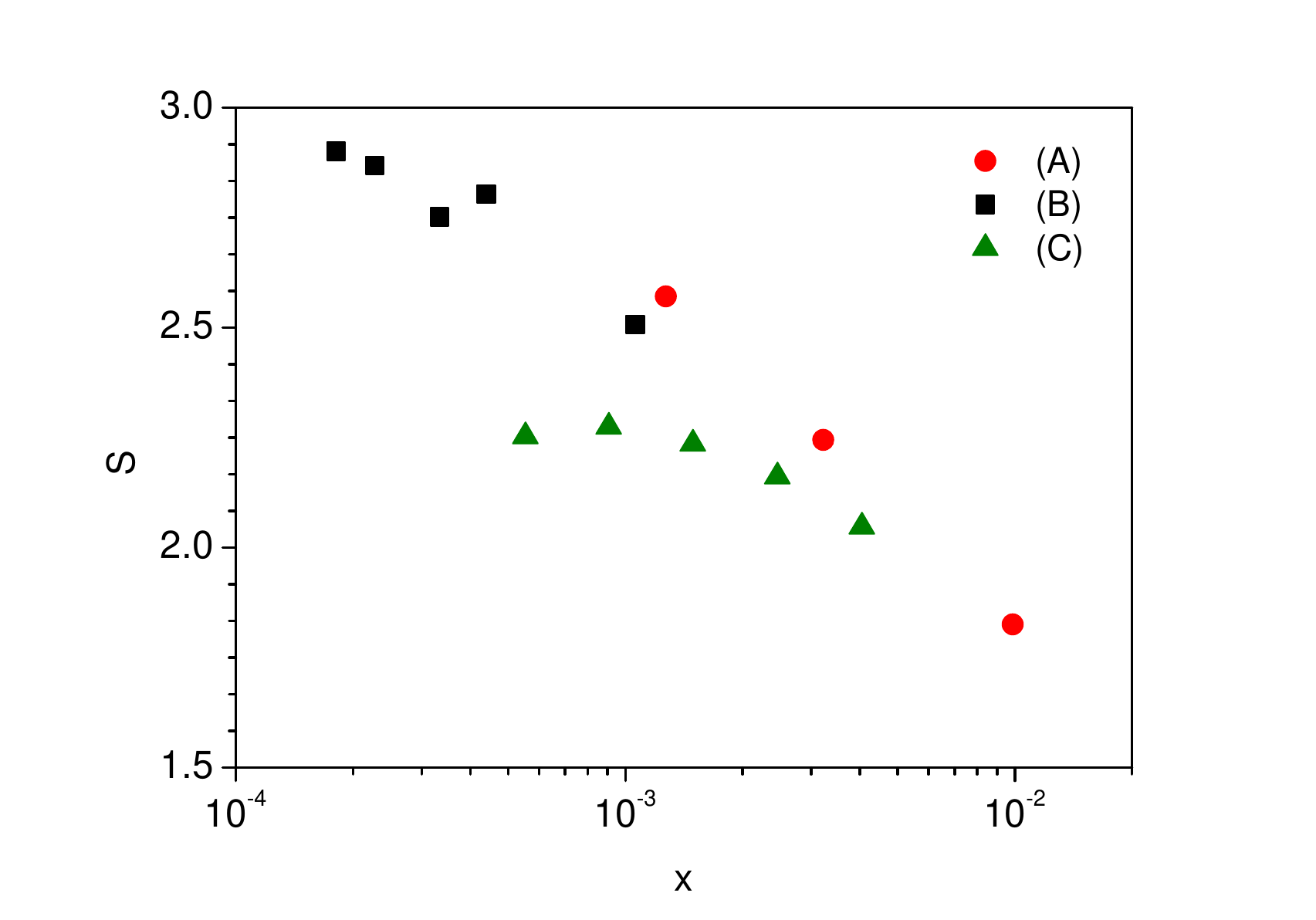}
\end{center}
\vspace{-5mm}
\caption{Hadron entropy derived from multiplicity distributions as a function of $x$. Symbols as in Fig.~\ref{SPD-1}.}
\label{SPD-2}
\end{figure}

\begin{figure}
\begin{center}
\includegraphics[scale=0.48]{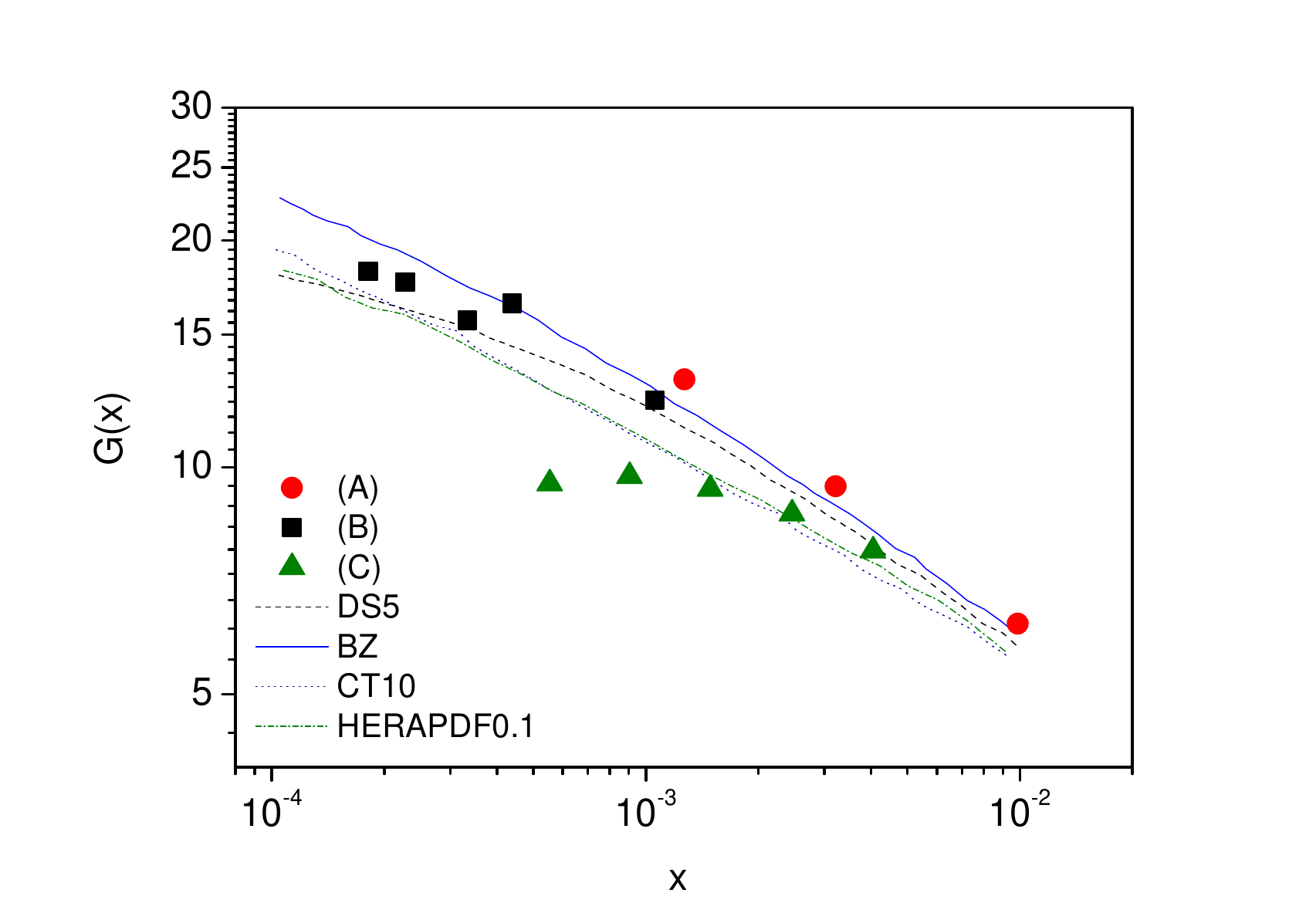}
\end{center}
\vspace{-5mm}
\caption{ Small-x behavior of gluon distribution, $G\left(x\right)=\exp\left(S\right)$, in comparison with different parametrizations. Dashed curve represent DS result, solid curve represent BZ model, dotted curve represent CT10, and dash-dotted curve represent HERAPDF0.1 result.  Symbols as in Fig.~\ref{SPD-1}.}
\label{SPD-3}
\end{figure}

Reference~\cite{Nair} demonstrated that the multiplicity distribution of inclusive photons - predominantly originating from $\pi^0$ decays - produced in inelastic proton-proton collisions at center-of-mass energies $\sqrt s = $ 900 GeV, 2.76 TeV, and 7 TeV, 
and measured in the forward hemisphere $2.3 < \eta < 3.9$  by the ALICE experiment at the LHC, is well described by the geometric-Poisson distribution. This distribution is a specific case of a compound Poisson distribution, suitable for modeling particle production in clustered systems: the number of clusters follows a Poisson distribution, while the number of particles within each cluster is governed by a geometric distribution. The extracted values of the distribution parameter $\Theta$, plotted 
as a function of the kinematic variable $x$, are presented in Fig.~\ref{SPD-1}. \footnote{For a given rapidity $y$, the variable $x$ is defined as $x=2 m_T \cosh(y)/\sqrt s$, where the transverse mass is given by $m_T=\sqrt{p_T^2+m^2}$.}

It is noteworthy that the multiplicity distribution arising from QCD parton cascades generally takes the form of the geometric distribution given by Eq.~(\ref{geometric}), where the parameter $\Theta=1/\left(\langle n\rangle+1\right)$ is directly related to the average parton multiplicity $\langle n\rangle$~\cite{PhysRevD.102.074008, PhysRevD.95.114008}. In the regime of small Feynman-x and large rapidity y, the parameter $\Theta$ exhibits an exponential dependence:
\begin{equation}
\Theta = \exp\left[-4\ln 2\bar{\alpha_S}y\right],
\end{equation}
where $\bar{\alpha_S} = \alpha_{S}N_c/\pi$ denotes the rescaled strong coupling constant, and $N_c$ is the number of QCD colors. Empirically, the extracted values of $\Theta$ correspond to relatively large values of the effective coupling: 
$\bar{\alpha_S} = $ 0.12, 0.16 and 0.19 for $\sqrt{s}$= 0.9, 2.76, and 7 TeV, respectively. This observation suggests that the hadronic final state may not be adequately described by a dilute system of partons. Instead, it may be more appropriate to treat the system as a dense partonic medium. In such a scenario, one expects a scaling relation of the form $1/\Theta = \langle n\rangle \sim Q_S^2 $, where $Q_S$ denotes the saturation scale~\cite{PhysRevD.102.074008}. Our analysis reveals that $Q_S^2$ exhibits a power-law dependence on the Feynman-x variable,
\begin{equation}
Q_S^2 \sim x^{-\Lambda}
\end{equation}
with an extracted exponent $\Lambda=0.31$. This result is in close agreement with previous determinations from deep inelastic scattering and diffractive processes, which report $\Lambda =$ 0.29 - 0.30~\cite{PhysRevD.59.014017, PhysRevD.60.114023}, as well as with theoretical predictions based on nonlinear QCD evolution~\cite{TRIANTAFYLLOPOULOS2003293}.

In the midrapidity region, charged-particle multiplicity distributions in proton-proton collisions have typically been described using the NBDs~\cite{ALICE}. For $\sqrt s =$ 8~TeV collisions, the ALICE Collaboration performed fits to the multiplicity distributions measured within various pseudorapidity windows—specifically, $|\eta|<2$,  $|\eta|<2.4$,  $|\eta|<3$,  $|\eta|<3.4$ and $-3.4<\eta<5$ —using a two-component NBD model~\cite{ALICE}. For the soft component, we extract the corresponding $\Theta$ parameter, shown in Fig.~\ref{SPD-1}. In contrast, the hard component exhibits a distinct behavior, suggesting that its dynamics are potentially governed by the quark structure function $x\Sigma (x)$, which does not exhibit the same scaling behavior as the gluon distribution.

Additionally, Fig.~\ref{SPD-1} displays the values of $\Theta$ extracted from the multiplicity distributions $P\left(N\right)$ of charged particles measured in the forward pseudorapidity region ($2<\eta<2.5$, $2.5<\eta<3$, $3<\eta<3.5$, $3.5<\eta<4$ and $4<\eta<4.5$) in  $\sqrt s =$ 7~TeV proton-proton collisions. In this kinematic regime, a single NBD provides an adequate description of the data.

The associated hadron entropy, calculated using Eq.~(\ref{S}), is presented in Fig.~\ref{SPD-2}. The obtained entropy values are found to be in qualitative agreement with the von Neumann entanglement entropy associated with gluon distributions, approximated by 
$\ln(n_{gluon})$~\cite{Tu}. In Fig.~\ref{SPD-3}, we compare our estimates of the gluon number density at fixed $x$, defined as $G(x)=\exp(S)$, with several widely used parton distribution function (PDF) parameterizations, including DS~\cite{DS} (dashed), BZ~\cite{BZ,BZ2} (solid), CT10~\cite{Guzzi,Liang} (dotted), and HERAPDF0.1~\cite{HERA,ZEUS} (dash-dotted). The comparison is performed at a fixed scale of $Q^2$=20 $GeV^2$; however, the dependence on $Q^2$ primarily affects the normalization rather than the shape of the distributions. This behavior is consistent with the Regge-like ansatz used to solve the linear Dokshitzer-Gribov-Lipatov-Altarelli-Parisi (DGLAP) evolution equation at small $x$, which leads to a factorized form:
\begin{equation}
G\left(x,Q^2\right)=G’\left(Q^2\right)G(x)= G’(Q^2)x^{-\Lambda},
\end{equation}
where $G’(Q^2)$ is a function of the hard scale $Q^2$, and $\Lambda$ represents the Regge intercept of the gluon distribution~\cite{DS}.

The HERAPDF0.1 parameterization is based on a combined analysis of H1 and ZEUS datasets, which resolves discrepancies between the two through cross-calibration. It plays an important role in making precise predictions for Standard Model processes. The CT10 PDFs, derived from global fits including HERA-1 neutral-current and charged-current DIS cross sections, represent one of the most comprehensive modern PDF sets. They are widely employed in phenomenological studies at the Tevatron, LHC, and other high-energy 
experiments. Our results exhibit a power-law behavior for the gluon density of the form $G(x)=1.5x^{-0.3}$, and the resulting shapes show good agreement with the trends predicted by the aforementioned parameterizations.

\section{Conclusions}
\label{Concl}

We highlight a compelling connection between hadronic observables and the underlying partonic dynamics involved in fragmentation and hadronization processes. Within the framework of compound multiplicity distributions for final-state hadrons, we identify a geometric distribution that effectively characterizes the gluon multiplicity emerging from QCD cascade evolution.

This study introduces novel tools for probing quantum information and complexity in QCD and opens the possibility for experimental investigations of quantum complexity through accessible hadronic observables. In particular, the entropy extracted from the hadron multiplicity distribution leads to a gluon density of the form $G\left(x\right)=\exp\left(S\right) \sim x^{-0.3}$, exhibiting small-x behavior consistent with predictions from various gluon density models.

We anticipate that these findings will motivate broader theoretical investigations and contribute to a deeper understanding of multiparticle production mechanisms in ultra-relativistic hadronic collisions. While the simplicity of this scenario is appealing, its applicability to the full range of experimental data remains to be confirmed through further studies based on dynamical modeling of the collision process.

\vspace*{0.3cm}
\centerline{\bf Acknowledgments}
\vspace*{0.3cm}
In preparation of this publication we used the resources of the Centre for Computation and Computer Modelling of the Faculty of Exact and Natural Sciences of the Jan Kochanowski University in Kielce, modernised from the funds of the Polish Ministry of Science and Higher Education in the “Regional Excellence Initiative” programme under the project RID/SP/00015/2024/01.

\appendix
\section{KNO scaling}
\label{KNO}

In high-energy particle physics, the Koba-Nielsen-Olesen (KNO) scaling, introduced over half a century ago, has played a foundational role in the empirical analysis of hadronic multiplicity distributions at high energies~\cite{Hegyi}. KNO scaling posits that 
the probability $P\left(N\right)$ of producing $N$ particles in a given collision process exhibits a universal scaling behavior of the form
\begin{equation}
\langle N\rangle P\left(N\right) = \psi\left(z\right),
\end{equation}
where $\langle N\rangle$ is the mean multiplicity and $z = N / \langle N\rangle$ is the scaling variable. Despite its empirical success, the derivation of KNO scaling directly from first principles in QCD remains nontrivial. In the context of perturbative QCD, theoretical studies suggest that KNO scaling naturally emerges in the asymptotic limit of high virtuality, particularly in jet fragmentation processes~\cite{DreminPhysRep}. The potential universality of KNO scaling and its connection to quantum entanglement entropy has also been discussed in Ref.~\cite{Liu}.

For small values of the geometric parameter $\Theta$ (corresponding to large average multiplicities $\langle n\rangle$), the approximation
\begin{equation}
\exp\left[-\Theta/\left(1-\Theta\right)\right] \simeq 1-\Theta
\end{equation}
holds, allowing Eq.~(\ref{geometric}) to be expressed as an exponential distribution: 
\begin{equation} 
P\left(n\right)=\frac{\Theta}{1-\Theta} \exp\left(-n \frac{\Theta}{1-\Theta} \right),
\label{A1} 
\end{equation}
Introducing the scaling variable $z=n/\langle n\rangle$, we obtain the KNO scaling form: 
\begin{equation} 
\langle n \rangle P\left(z\right) = \exp\left(-z\right). 
\label{A2} 
\end{equation}

Considering a total multiplicity $N=\sum_i^k n_i$ arising from $k$ independent sources, each contribution a multiplicity $n_i$ drawn from the exponential distribution~(\ref{A2}), the total mean multiplicity becomes $\langle N\rangle = k\langle n \rangle$. The sum of $k$ exponentially distributed random variables is known to follow a gamma distribution, leading to the following KNO-scaled form:
\begin{equation} 
\langle N \rangle P\left(N\right) = \frac{k^k}{\Gamma\left(k\right)} z^{k-1} \exp\left(-kz\right), 
\label{A3} 
\end{equation} 
where $z=N/\langle N\rangle$.

Notably, the scaling form in Eq.~(\ref{A3}) has also been derived by various other authors, though their theoretical motivations differ from the approach presented here \cite{SM}.

\section{$P(N)$ of the compound geometric distributions}
\label{PN}

The probability distribution $P\left(N\right)$ of the random variable $N$ is given by the derivatives of generating function $H\left(z\right)$,
\begin{equation}
P\left(N\right)=\frac{1}{N!}\left.\frac{\partial ^N H\left(z\right)}{\partial z^N}\right|_{z=0}.
\label{B0}
\end{equation}
Generating function~(\ref{GNBD}) leads to well known negative binomial distribution
\begin{equation}
P\left(N\right)=\binom{N+k-1}{N} \left(1-\Theta'\right)^N\left(\Theta'\right)^k.  
\label{B1}
\end{equation}
For generating function~(\ref{GGB}) we have geometric-binomial distribution
\begin{equation}
\begin{split}
P\left(0\right)&=\left(1-\frac{\lambda}{K}+\frac{\lambda}{K}\Theta' \right)^K \\
 P\left(N\right)&= \left(1-\Theta' \right)^N  \sum_{m = 1}^{K} \binom{K}{m} \binom{N+m-1}{N}  \left(\frac{\lambda}{K} \Theta' \right)^m  \left(1-\frac{\lambda}{K}\right)^{K-m}, 
\end{split}
\label{B2}
\end{equation}
and for generating function~(\ref{GGP}) we obtain geometric-Poisson distribution
\begin{equation}
\begin{split}
P\left(0\right)&=\exp(-\lambda)\\
 P\left(N\right)&= \sum_{m = 1}^{N} \frac{e^{-\lambda}\lambda^m}{m!} \binom{N - 1}{m - 1} \Theta^{m} \left(1 - \Theta\right)^{N -m}. 
\end{split}
\label{B3}
\end{equation}

\end{document}